# Insulator, semiclassical oscillations and quantum Hall liquids at low magnetic fields

Shun-Tsung Lo[1], Yi-Ting Wang[2], G Bohra[3], E Comfort[3], T-Y Lin[3], M-G Kang[3], G Strasser[3], J P Bird[3], C F Huang[4], Li-Hung Lin[5], J C Chen[6], and C-T Liang[1,2]

[1]*Graduate Institute of Applied Physics, National Taiwan University, Taipei 106, Taiwan*

[2]*Department of Physics, National Taiwan University, Taipei 106, Taiwan*

[3]*Department of Electrical Engineering, University at Buffalo, The State University of New York, Buffalo, New York 14260-1920*

[4]*National Measurement Laboratory, Center for Measurement Standards, Industrial Technology Research Institute, Hsinchu 300, Taiwan*

[5]*Department of Electrophysics, National Chiayi University, Chiayi 600, Taiwan*

[6]*Department of Physics, National Tsinghwa University, Hsinchu 300, Taiwan*

e-mail: jbird@buffalo.edu and ctliang@phys.ntu.edu.tw

## Abstract

Magneto-transport measurements are performed on two-dimensional GaAs electron systems to probe the quantum Hall (QH) effect at low magnetic fields. Oscillations following the Shubnikov-de Haas (SdH) formula are observed in the transition from the insulator to QH liquid when the observed almost temperature-independent Hall slope indicates insignificant interaction correction. Our study shows that the existence of SdH oscillations in such a transition can be understood based on the non-interacting model.

**I. Introduction**

For a two-dimensional electron gas (2DEG) subject to a perpendicular magnetic field $B$, its energy can only take discrete values governed by Landau quantization $E_n = (n+1/2)\, \hbar eB/m^*$. Here $n$ denotes a non-negative integer, $\hbar$ and $e$ are the reduced Planck constant and electron charge, and $m^*$ represents the effective mass. Landau quantization plays an important role in determining the magnetotransport properties of a two-dimensional (2D) system. The Shubnikov-de Haas (SdH) formula [1-4], which describes the oscillating amplitude $\Delta\rho_{SdH}$ in the longitudinal resistivity $\rho_{xx}$ as a result of such quantization at low $B$, is given by

$$\Delta\rho_{SdH}(B,T) = 4\rho_0 \exp(-\pi/\mu_q B) D(B,T). \qquad (1)$$

Here $T$ represents the temperature, the parameter $\rho_0$ is expected to be close to the value of $\rho_{xx}$ at $B=0$ (while there are reports on deviations from this [5]), $\mu_q$ corresponds to the quantum mobility, and the Lifshitz-Kosevich (LK) factor

$$D(B,T) = \frac{2\pi^2 k_B m^* T / \hbar eB}{\sinh(2\pi^2 k_B m^* T / \hbar eB)}, \qquad (2)$$

with $k_B$ as the Boltzmann constant. Equations (1) and (2) can be used to determine the effective mass and quantum mobility and one can obtain the carrier concentration $n$ from the oscillating period in $1/B$. These parameters are important physical quantities not only in the field of condensed matter physics, but also in materials science and device engineering.

It is worth mentioning that the SdH formula can be derived by considering a Fermi-liquid metal where the semiclassical approach is valid because of unimportant quantum localization effects [6]. With increasing $B$, however, we must incorporate such effects to understand the appearance of the quantum Hall effect (QHE) under Landau quantization [7]. Based on the floating-up picture, Kivelson, Lee and Zhang [8] proposed the global phase diagram (GPD) for the QHE. Within the GPD, the 2D system at low $B$ is an insulator denoted by the number "0" rather than a Fermi-liquid metal. To enter the integer quantum Hall effect from the insulator, the system must pass through the quantum Hall (QH) state of $\nu=1$, where $\nu$ denotes the Landau-level filling factor. When one takes the electron spin degree of freedom into consideration, the 2D GaAs electron system can enter the $\nu=2$ QH state directly from the insulator when the spin-splitting is unresolved [9, 10]. Therefore, the only allowed insulator (I)-QH transitions are the 0-1 and 0-2 transitions, for the 2D GaAs electron systems to enter $\nu=1$ or 2 QH states after leaving the insulator. But the direct I-QH transition, where the 2D system enters QH states of $\nu>2$ directly from the insulator, has been observed [11-17]. Actually the GPD is for an ideal 2D system where metallic regimes are unstable based on suitable assumptions, which may become invalid experimentally at low $B$. Huckestein [18] has suggested that, with increasing $B$, the direct I-QH transition corresponds to the onset of Landau quantization at $B=1/\mu$, below which the standard QH theory is invalid. Here $\mu$ is the mobility. Since it is possible to define different mobilities, $\mu$ should be the quantum mobility $\mu = \mu_q$ such that the product $\mu B$ equals the ratio of the Landau-level broadening to level spacing. However, it has been pointed out that such a transition can occur when $B<1/\mu_q$ [16, 17], and the transition point should be at the magnetic field $B=1/\mu_c$ rather than $1/\mu_q$ at low enough $T$ when the ballistic-like correction is negligible [19]. Here $\mu_c$ denotes the classical mobility. It has been pointed out that the insulating regime in the observed direct I-Q transition may not be a genuine insulating phase/state due to finite size/temperature effect [18]. Recent studies reveal that the semiclassical SdH oscillations can survive under the mobility gap resulting from quantum localization [20], and more studies are needed to clarify the 2D transport properties under Landau quantization at low magnetic fields.

In Refs. [15] and [16] reported by our group, magnetoresistance oscillations following the semiclassical SdH formula (Eq. 1) were observed when GaAs 2D electron systems undergo an I-QH transition from the low-field insulator, directly to QH states of $\nu>2$. Therefore, metallic behavior can survive when there exists a localization-induced insulator or QH liquid. In these reports, the quantum correction due to the electron-electron (e-e) interaction exists, as evidenced by the logarithmic $T$ dependence of the Hall slope. While such an interaction correction may induce metallic behavior [21-24], the metallic regime may also appear when the effective size is smaller than the scale for strong localization [8]. Nita, Aldea and Zittartz [25] predicted the existence of the metallic regime in the direct I-QH transition when the magnetic field results in delocalization in a finite-size system [26]. In such a regime, Landau quantization can result in magneto-oscillations without inducing QH states. It is worth pointing out that the samples reported in Ref. [15] and Ref. [16] featured either delta-doping or had InAs quantum dots inserted. Consequently, it is of interest to study the corresponding behavior for a typical modulation-doped GaAs/AlGaAs heterostructure. Moreover, to further clarify size effects and the role of interaction correction in the low-field Landau quantization and direct I-QH transition, it is of fundamental interest to probe 2D systems with negligible e-e interaction. In the presence of a direct I-QH transition, we note that the SdH formula does not extend to the QH side in Ref. [15], while such a formula could not be investigated on the insulating side in Ref. [16]. More studies are required to clarify whether oscillations following Eq. (1) can span from the insulator to the QH regime.

In this paper, we report observations of SdH oscillations following Eq. (2) in the low-field insulator. While in our previous study [15] these oscillations are observed under the significant correction resulting from e-e interaction, in this report such oscillations and insulator-like behavior coexist when the almost $T$-independent Hall slope indicates the weak interaction strength. Therefore, our observations reveal that the appearance of SdH oscillations in the insulator-like regime is irrelevant to the strength of the interaction correction and can be understood based on the non-interacting model. In addition, oscillations following the SdH formula are found to cover both sides of direct I-QH transitions just as predicted by Nita, Aldea, and Zittartz [25]. For convenience, in the following we describe the experimental results on magnetoresistance oscillations and the direct I-QH transition in section II. A discussion of the experimental results is given in Sec. III, and our conclusions are provided in section IV.

**II. Semiclassical magneto-oscillations and direct insulator-quantum Hall transition**

Two devices, Sample A and Sample B were studied in this work. For sample A, the following layer sequence was grown on a semi-insulating GaAs (100) substrate by molecular beam epitaxy (MBE): 50 nm GaAs, 30 periods of a 2-nm AlAs/2-nm GaAs superlattice, 1 μm GaAs, 20 nm $Al_{0.33}Ga_{0.67}As$, with a $3 \times 10^{11}$ cm$^{-2}$ Si delta-doping layer, 8 periods of 2.05 nm $Al_{0.33}Ga_{0.67}As$/0.9 nm AlAs/2.05 nm $Al_{0.33}Ga_{0.67}As$, and a 5 nm GaAs cap layer. For sample B, the MBE layer sequence on a GaAs (100) SI substrate was as follows: 30 nm GaAs, 30 periods of a 2-nm AlAs/2-nm GaAs superlattice, 1 μm GaAs, 20 nm $Al_{0.33}Ga_{0.67}As$, A Si-doping layer with a concentration of $10^{18}$ cm$^{-3}$, 40 nm $Al_{0.33}Ga_{0.67}As$, and a 5 nm GaAs cap layer. The devices were made with a Hall pattern by standard wet-etching processes and optical lithography. AuGeNi alloy was evaporated and annealed to form Ohmic contacts to the 2DEG. For sample A, Au/Ti was evaporated to form a front gate on top of the Hall bar, whereas Sample B was an ungated device.

Figure 1 shows curves of $\rho_{xx}$ and $\rho_{xy}$ for sample A with a voltage of -0.075 V applied on the front gate. There is a well-developed quantum Hall state with filling factor ν=2 for $B$~3-3.8 T in Fig. 1, and we can see from the left inset that the sample becomes an insulator as $B<0.66$ T≡$B_c$, in the sense that $\rho_{xx}$ increases with decreasing $T$. The blue points in the right inset to Fig. 1 shows the $T$-dependence of the slope $R_H(T)$ of $\rho_{xy}$ at low $B$, and the regression line of $R_H(T)$-ln$T$ is denoted by the blue dashed one. The temperature dependences of $\rho_{xx}$ are different on the two sides of $B_c$, near which there is no plateau of ν=2 or 1, and the slope of $\rho_{xy}$ in the inset (b) is almost $T$-independent within experimental error. Therefore, the direct I-QH transition is observed [27], and the almost $T$-independent Hall slope as shown in inset (b) to Fig. 1 indicates that the correction due to the e-e interaction is very weak [28]. From the period of SdH oscillations in $1/B$, the carrier concentration $n = 1.67 \times 10^{11}$ cm$^{-2}$. By analyzing the low-field amplitude in $\rho_{xx}$, as shown in Fig. 2 (a), the experimental data $\ln[\Delta\rho_{xx}/D(B, T)]$ as a function of $1/B$ show a good fit to Eq. (1) when we take $m^*$ as the expected value 0.067 $m_0$. We can obtain the quantum mobility $\mu_q = (0.665 \pm 0.008)$ m$^2$/Vs, and its inverse $1/\mu_q \cong 1.5$ T, much larger than $B_c = 0.66$ T. Therefore, the direct I-QH transition can occur as $B < 1/\mu_q$ even when there is no significant interaction correction.

To further study the direct I-QH transition, together with magnetoresistivity oscillations in Sample A, we vary the applied gate voltage $V_g$. Figure 1 (b) shows the curves of $\rho_{xx}$ and $\rho_{xy}$ for sample A at $V_g = 0$, under which we have the carrier concentration $n = 2.33 \times 10^{11}$ cm$^{-2}$ from the period of SdH oscillations in $1/B$. As shown in the inset to Fig. 3, at $V_g = 0$ we can see a magnetoresistance bump resulting from Landau quantization in the insulating regime although the almost $T$-independent Hall slope as shown by the red one in inset (b) to Fig. 1 indicates insignificant e-e interaction correction [21-24]. Therefore, our study reveals that Landau quantization can result in oscillations in the low-field insulator when the e-e interaction correction is weak. In addition, together with the magneto-oscillations on the quantum Hall side, such a bump supports that the metallic regime characterized by Landau quantization can cover both sides of the direct I-QH transition just as what is predicted by Nita, Aldea, and Zittartz [25]. We can see from Fig. 2 (b) that the oscillating amplitude, including the bump in the low-field insulator, at $V_g = 0$ fits Eq. (1) well with $\mu_q = (0.645 \pm 0.006)$ m$^2$/Vs when $B < 1.7$ T. So, the direct I-QH transition occurs in the metallic regime where Landau quantization induces oscillations following the semiclassical SdH formula. The data presented in Fig. 2 (b) reveals that the magnetoresistance oscillations in the insulating regime and the QH-like region can be fitted to Eq. (1). However, there are only a couple of data points in the insulating regime. In order to test whether it is possible to observe even more oscillations in the insulating regime, magneto-transport measurements are performed on sample B, where the Hall slope is also almost $T$-independent as shown in the inset to Fig. 3 and thus e-e interaction correction is weak. The carrier concentration $n = 3.6 \times 10^{11}$ cm$^{-2}$ from period of SdH oscillations in $1/B$. Figure 3 shows the curves of the longitudinal resistance $R_{xx}$ in Sample B. The direct I-QH transition occurs at $B_c = 1.4$ T, below which the sample is an insulator because $R_{xx}$ increases with decreasing $T$. Figure 2 (c) shows the logarithmic of amplitude divided by $D(B,T)$ $\ln[\Delta\rho_{xx}/D(B, T)]$ as function of $1/B$ at $T = 2.1$ K for Sample B. We can see that the measured $\ln[\Delta\rho_{xx}/D(B, T)]$ shows a good linear dependence on $1/B$ over at low temperatures. From the fit to Eq. (1), the determined quantum mobility is $(0.789\pm0.016)$ m$^2$/Vs. Most importantly, we can see that there is a good fit spanning from the insulator to the QH-like regime. Therefore we have presented compelling experimental evidence that magneto-oscillations governed by Eq. (1) can indeed cover from the insulator to the QH-like regime.

**III Discussion**

It has been suggested by Huckestein [18] that the direct I-QH transition can be identified as the crossover from weak localization to the onset of Landau quantization as $B=1/\mu$, and the standard QH theory is valid in the QH regime as $B>1/\mu$. Just as mentioned above, here $\mu$ should be the quantum mobility $\mu_q$, a measure of the ratio of the Landau-level broadening to the level spacing. However, the magnetoresistance oscillations in the insulating regime revealed that the onset of Landau quantization does not correspond to such a transition [15], and it is well-known that these oscillations can follow Eq. (1) when $B<1/\mu_q$ [6]. Our previous study [14, 15] showed that the crossover from weak localization to Landau quantization can cover a wide range of magnetic field rather than the single point corresponding to the direct I-QH transition. While the onset of Landau quantization does not occur as $B=1/\mu_q$, in our previous report [16] the deviation from the conventional SdH theory occur as $B\sim1/\mu_q$ and the scaling behaviors expected in the standard QH theory appear as $B>1/\mu_q$. Therefore, the magnetic field $B=1/\mu_q$ may separate a low-$B$ regime, following semiclassical transport, from the high-$B$ liquid governed by the standard QH theory, in some samples. On the other hand, in sample B Eq. (1) remains valid when $B>1/\mu_q$ just as in high-mobility samples, where the semiclassical metallic transport and quantum localization behaviors can coexist [20]. The positive magneto-resistance background when $B>1.4$ T, as shown in Fig. 4, may be the reason why oscillations at $B>1/\mu_q$ can be well approximated by the conventional SdH formalism [29]. Very recently observations of the direct I-QH transition in high-density InGaAs-based 2DEG [30] revealed that percolation may play a role, and it has been reported that $\rho_{xx}\sim\rho_{xy}$ can occur as $B<B_c$ [30]. Furthermore, the direct I-QH transition is not always dominated by weak localization effect [30]. Therefore further studies are required to obtain a better understanding of the low-field Landau quantization.

It has been pointed out that the critical longitudinal and Hall resistivities are approximately the same at the direct I-QH transition, and thus such a transition occurs at $\rho_{xx}\sim\rho_{xy}$ [11]. On the other hand, we can see from Ref. [31] that $\rho_{xx}$ can be larger than $\rho_{xy}$ at the I-QH transition even as the filling factor $\nu>1$. In our study, the direct I-QH transition does not occur at $\rho_{xx}\sim\rho_{xy}$. Therefore our experimental results, together with existing reports [15, 30], suggest that the relation $\rho_{xx} \sim \rho_{xy}$ may fail whilst such a possible universality may hold in some samples. We note that universalities expected in high-field quantum Hall transitions may also become invalid [32-34]. For an example, in the seminal work of Li *et al.* [34], it has been pointed out that the scaling exponent can depend experimentally on the nature of disorder. To compare the experimental and theoretical results, it is important to clarify the difference between the real systems and theoretical models. In our system, *in-situ* tilted field measurements reveal that the low-field insulator is not a perfect 2D phenomenon while most quantum Hall theories are for ideal 2D systems. Further studies are required to obtain a thorough understanding of the expected universalities.

For non-interacting electrons with negligible interference effects, at low $T$ the longitudinal and Hall conductivities are given by

$\sigma_{xx} = ne\mu_c/(1+\mu_c^2 B^2)$,     (3)

$\sigma_{xy} = ne\mu_c^2 B/(1+\mu_c^2 B^2)$,     (4)

based on the Drude model [35]. When the e-e interaction effect is taken into account, in the diffusive regime the correction term [23, 36]

$\delta\sigma_{xx}(T) = \dfrac{e^2}{\pi h} K_{ee} \ln T + \delta_0$     (5)

has been introduced to modify Eq. (3) while the validity of Eq. (4) is still expected. Here $K_{ee}$ is the interaction coefficient, and $\delta_0$ denote the $T$-independent factor [37]. Although $\delta_0$ can be of a different form, it has been shown that the equation $\delta\sigma_{xx}(T_2) - \delta\sigma_{xx}(T_1) = \dfrac{e^2}{\pi h} K_{ee} \ln T_2/T_1$ [37] provides a good approximation for the difference at any two temperatures $T_1$ and $T_2$. Therefore, we can choose a reference temperature $T_r$ and replace Eq. (5) by

$\delta\sigma_{xx,r}(T) = \dfrac{e^2}{\pi h} K_{ee} \ln T/T_r$     (6)

to estimate the $T$ dependence of $\sigma_{xx}$. Then the longitudinal resistivity and Hall slope $R_H(T)$ are

$\rho_{xx}(B,T) = \rho_{xx}(B,T_r) - \dfrac{1}{(ne\mu_c)^2}(1-\mu_c^2 B^2)\dfrac{K_{ee} e^2}{\pi h} \ln T/T_r$     (7)

$R_H(T)/R_H(T_r) = 1 - \dfrac{2K_{ee}}{\pi h n^2 \mu_c s_H(T_r)} \ln T/T_r$,     (8)

respectively. From Eq. (7), we can obtain the classical mobility $\mu_c = 1/B_c$ because the transition point to separate the insulator and quantum Hall liquid is at the magnetic field $B_c = 1/\mu_c$ rather than $1/\mu_q$. On the other hand, Eq. (8) provides the $T$-dependence of the Hall slope due to the interaction correction. We can compare the strengths of the interaction corrections in different 2DEGs by checking $R_H(T)/R_H(T_r)$ based on Eq. (8).

In this report, just as mentioned above, SdH oscillations are observed together with the direct I-QH transition as the Hall slope is almost $T$-independent. On the other hand, in Ref. 15 they are observed as the significant $T$-dependence of $R_H$ indicates the importance of the interaction correction to such transition. Figure 5 compares $R_H(T)/R_H(T_r)$ in this study to that in Ref. 15 as $V_g$=-0.07 V, and the red, blue, black, and green points represent such ratios for sample A at $V_g$=0 V, sample A at $V_g$=-0.075 V, sample B, and the sample discussed in Ref. 15, respectively. Here we take $T_r$=2.05 K and estimate $R_H(T_r)$ from the regression line of the $R_H(T)$-ln$T$ at $T_r$. We can see from Fig. 5 that the $T$-dependences of $R_H(T)$ in this study are much weaker than that in our previous study while SdH oscillations appear near $B_c$ in all cases. Therefore, the appearance of SdH oscillations near the direct I-QH transition is irrelevant to the strength of the interaction correction and can be understood based on the non-interacting model.

In Fig. 5, the red, blue, black, and green lines denote $R_H(T)/R_H(T_r)$ expected under Eq. (8) for sample A at $V_g$=0 V, sample A at $V_g$=-0.075 V, sample B, and the sample discussed in Ref. 15 for $V_g$=-0.07 V, respectively. The parameters $n$ and $\mu_c$ in Eq. (8) are obtained from the period of SdH oscillations in $1/B$ and the inverse of the transition magnetic field $B_c$, respectively. While the experimental values fit the expected line very well for the sample described in Ref. [15], in this study the $T$-dependence of $R_H(T)/R_H(T_r)$ is so weak that such ratios do not follow Eq. (8). In our work, the interaction correction term is well below the predicted value. The experimental errors on the Hall slopes are smaller than 0.1 % while the predicted changes are larger than 3 % for both samples A and B, so the deviations to the theoretical calculations are not due to the measurement uncertainties. In Ref. [34], the measured e-e interaction strength is also lower than that predicted by Zala, Narozhny and Aleiner [21]. We note that the theory only considers short-range scattering [34], but scatterings of different ranges may coexist within the same disordered 2DEG. In addition, the $B$-dependence of $\rho_{xx}$ can deviate from that expected based on Eq. (7), which indicates the importance of mechanisms beyond the considered interaction correction [16]. Within the Drude model, $\mu_c = 1/B_c$ is expected to be close to $\mu_q$. For sample B, $\mu_c$=0.71 m$^2$/Vs is close to $\mu_q$ =0.79 m$^2$/Vs. However, for Sample A, $\mu_c$=1.49 m$^2$/Vs is substantially different from $\mu_q$~0.63 m$^2$/Vs, consistent with existing experimental results [14, 38–41]. The discrepancy between theoretical model and experimental results suggests that more studies are required for obtaining a thorough understanding of interactions in disordered systems. On the other hand, Eqs. (1) and (2) hold true no matter whether Eq. (7) is valid for the Hall slope $R_H$, which also indicates that the validity of SdH formula is irrelevant to the interaction correction and thus can be understood based on the non-interacting model.

## IV Conclusion

In summary, we have presented detailed magentotransport measurements on two GaAs-based 2DEGs. We have shown that magnetoresistance oscillations can span from the insulating to QH-like regions when e-e interaction correction is not significant. Moreover, a single mobility cannot be used to describe our experimental results.

## Acknowledgments

This work was funded by the Department of Energy, USA. L.H.L., J.C.C. and C.T.L. acknowledge financial support from the NSC, Taiwan. C.T.L. would like to thank National Taiwan University for providing financial support (grant no: 98R0704 and grant no: 10R809083B.). We would like to thank Ming-Yang Li, Kuan-Ting Lin, Yiping Lin, and Dong-Sheng Luo at NTHU, and Shih-Wei Lin and S.-D. Lin at NCTU for experimental help.

**References**

[1] Fowler A B, Fang F F, Howard W E and Stiles P J 1966 *Phys. Rev. Lett.* **16** 901

[2] Ando T 1974 *J. Phys. Soc. Jpn.* **37** 1233

[3] Ando T, Matsumoto Y and Uemura Y 1975 *J. Phys. Soc. Jpn.* **39** 279

[4] Isihara A and Smrcka L 1986 *J. Phys. C* **19** 6777

[5] Chen J H, Hang D R, Huang C F, Huang T-Y, Lin J Y, Lo S H, Hsiao J C, Lin M-G, Simmons M Y, Ritchie D A and Liang C-T 2007 *J. Korean Phys. Soc.* **50** 776

[6] For example, see Martin G W, Maslov D L and Reizer M Y 2003 *Phys. Rev. B* **68** 241309 and references therein

[7] von Klitzing K, Dorda G and Pepper M 1980 *Phys. Rev. Lett.* **45** 494

[8] Kivelson S, Lee D-H and Zhang S-C 1992 *Phys. Rev. B* **46** 2223

[9] Jiang H W, Johnson C E, Wang K L and Hannahs S T 1993 *Phys. Rev. Lett.* **71** 1439

[10] Hughes R J F, Nicholls J T, Frost J E F, Linfield E H, Pepper M, Ford C J B, Ritchie D A, Jones G A C, Kogan E and Kaveh M 1994 *J. Phys.: Condens. Matter* **6** 4763

[11] Song S H, Shahar D, Tsui D C, Xie Y H and Monroe D 1997 *Phys. Rev. Lett.* **78** 2200

[12] Lee C H, Chang Y H, Suen Y W and Lin H H 1997 *Phys. Rev. B* **56** 15238

[13] Smorchkova I P, Samarth N, Kikkawa J M and Awschalom D D 1998 *Phys. Rev. B* **58** R4238

[14] Huang T-Y, Juang J R, Huang C F, Kim G-H, Huang C-P, Liang C-T, Chang Y H, Chen Y F, Lee Y and Ritchie D A 2004 *Physica E* **22** 240

[15] Huang T-Y, Liang C-T, Kim G-H, Huang C F, Huang C-P, Lin J-Y, Goan H-S and Ritchie D A 2008 *Phys. Rev. B* **78** 113305

[16] Chen K Y, Chang Y H, Liang C-T, Aoki N, Ochiai Y, Huang C F, Li-Hung L, Cheng K A, Cheng H H, Lin H H, Jau-Yang W and Sheng-Di L 2008 *J. Phys.: Condens. Matter* **20** 295223

[17] Lo S-T, Chen K Y, Lin T L, Lin L-H, Luo D-S, Ochiai Y, Aoki N, Wang Y-T, Peng Z F, Lin Y, Chen J C, Lin S-D, Huang C F and Liang C-T 2010 *Solid State Commun.* **150** 1902

[18] Huckestein B 2000 *Phys. Rev. Lett.* **84** 3141

[19] The classical mobility can be taken as the renormalized mobilty when the temperaure approaches zero. See Minkov G M, Germanenko A V, Rut O E, Sherstobitov A A, Larionova V A, Bakarov A K and Zvonkov B N 2006 *Phys. Rev. B* **74** 045314

[20] Hang D R, Huang C F and Cheng K A 2009 *Phys. Rev. B* **80** 085312
[21] Dobrosavljević V, Abrahams E, Miranda E and Chakravarty S 1997 *Phys. Rev. Lett.* **79** 455
[22] Dubi Y, Meir Y and Avishai Y 2005 *Phys. Rev. Lett.* **94** 156406
[23] Zala G, Narozhny B N and Aleiner I L 2001 *Phys. Rev. B* **64** 214204
[24] Clarke W R, Yasin C E, Hamilton A R, Micolich A P, Simmons M Y, Muraki K, Hirayama Y, Pepper M and Ritchie D A 2008 *Nat. Phys.* **4** 55
[25] Nita M, Aldea A and Zittartz J 2000 *Phys. Rev. B* **62** 15367
[26] Nita M 2010 private communications
[27] Huang C F, Chang Y H, Lee C H, Chou H T, Yeh H D, Liang C-T, Chen Y F, Lin H H, Cheng H H and Hwang G J 2002 *Phys. Rev. B* **65** 045303
[28] For example, see Simmons M Y, Hamilton A R, Pepper M, Linfield E H, Rose P D and Ritchie D A 2000 *Phys. Rev. Lett.* **84** 2489 and references therein
[29] Hang D R, Huang C F, Zhang Y W, Yeh H D, Hsiao J C and Pang H L 2007 *Solid State Commun.* **141** 17
[30] Gao K H, Yu G, Zhou Y, Wei L, Lin T, Shang L Y, Sun L, Yang R, Zhou W Z, Dai N, Chu J H, Austing D G, Gu Y and Zhang Y G 2010 *J. Appl. Phys.* **108** 063701
[31] Hilke M, Shahar D, Song S H, Tsui D C and Xie Y H 2000 *Phys. Rev. B* **62** 6940
[32] Koch S, Haug R J, Klitzing K v and Ploog K 1991 *Phys. Rev. B* **43** 6828
[33] Huang C F, Chang Y H, Cheng H H, Yang Z P, Wang S Y, Yeh H D, Chou H T, Lee C P and Hwang G J 2003 *Solid State Commun.* **126** 197
[34] Li W, Vicente C L, Xia J S, Pan W, Tsui D C, Pfeiffer L N and West K W 2009 *Phys. Rev. Lett.* **102** 216801
[35] Murzin S 2010 *JETP Lett.* **91** 155
[36] Minkov G M, Rut O E, Germanenko A V, Sherstobitov A A, Shashkin V I, Khrykin O I and Zvonkov B N 2003 *Phys. Rev. B* **67** 205306
[37] Goh K E J, Simmons M Y and Hamilton A R 2008 *Phys. Rev. B* **77** 235410
[38] Cho H-I, Gusev G M, Kvon Z D, Renard V T, Lee J-H and Portal J C 2005 *Phys. Rev. B* **71** 245323
[39] Minkov G M, Rut O E, Germanenko A V, Sherstobitov A A, Zvonkov B N, Uskova E A and Birukov A A 2002 *Phys. Rev. B* **65** 235322
[40] Minkov G M, Germanenko A V, Rut O E, Sherstobitov A A and Zvonkov B N 2010 *Phys. Rev. B* **82** 035306
[41] Huang T Y, Liang C-T, Kim G H, Huang C F, Huang C P and Ritchie D A 2010 *Physica E* **42** 1142

Figure Captions

Figure 1 (a) Longitudinal magnetoresistivity $\rho_{xx}(B)$ and Hall resistivity $\rho_{xy}(B)$ at various temperatures when a front gate voltage of -0.075 V is applied on Sample A. The left inset is a zoom-in of the low-field $\rho_{xx}(B)$. The right Inset shows the measured Hall slopes $R_H$ versus the logarithm of the measurement temperature ln$T$ for $V_g$=-0.075 V and $V_g$=0 V. The dashed lines correspond to the best linear fits to the data. (b) $\rho_{xx}(B)$ and $\rho_{xy}(B)$ at various temperatures when zero gate voltage is applied on Sample A. The inset is a zoom-in of the low-field $\rho_{xx}(B)$ showing a SdH bump in the insulating region.

Figure 2 ln[$\Delta\rho_{xx}/D(B, T)$] as function of 1/$B$ for (a) $V_g$=-0.075 V and (b) $V_g$=0 V (Sample A), and (c) Sample B. The straight lines correspond to a fit to Eq. (1).

Figure 3 Longitudinal magnetoresistance $R_{xx}(B)$ and Hall resistivity $\rho_{xy}(B)$ at various temperatures for Sample B. Inset shows the measured Hall slopes $R_H$ versus the logarithm of the measurement temperature ln$T$. The dashed line corresponds to the best linear fit to the data.

Figure 4 $R_{xx}$ and the average of the fitted envelope functions (in dashed line) going through the maxima (in dotted line) and minima (in dot-dash line) of the oscillations as a function of $B$ (Sample B).

Figure 5 The ratios of the Hall slopes to those at a reference temperature $T_r$=2.05 K $R_H(T)/R_H(T_r)$ as a function of ln$T$. The red, blue, black, and green symbols represent experimental results taken from sample A at the gate voltage $V_g$=0 V, sample A at $V_g$=-0.075 V, sample B, and the sample discussed in Ref. 15 for $V_g$=-0.07 V. For each data set, the corresponding Hall slope predicted by Eq. (8) is given by the line of the same color.

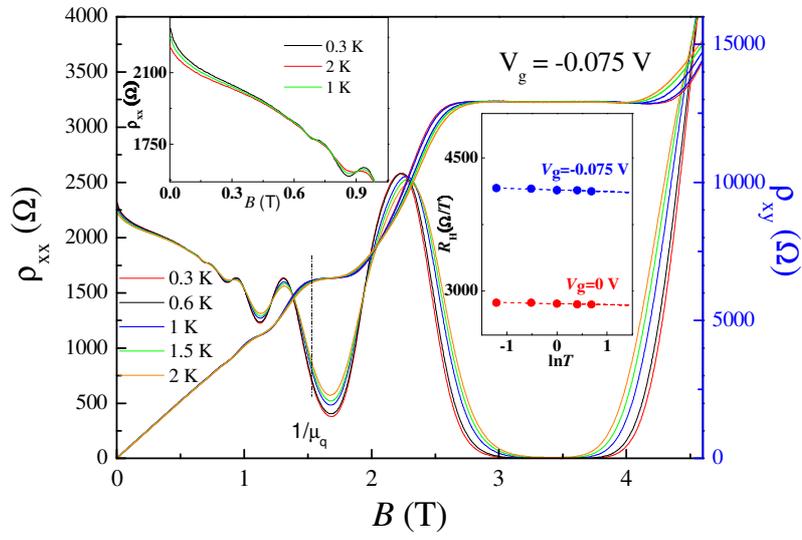

Figure 1 (a)

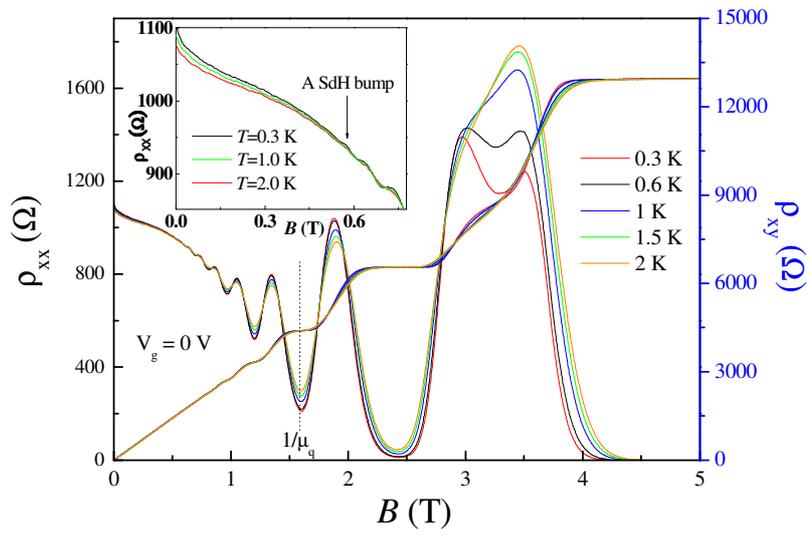

Figure 1 (b)

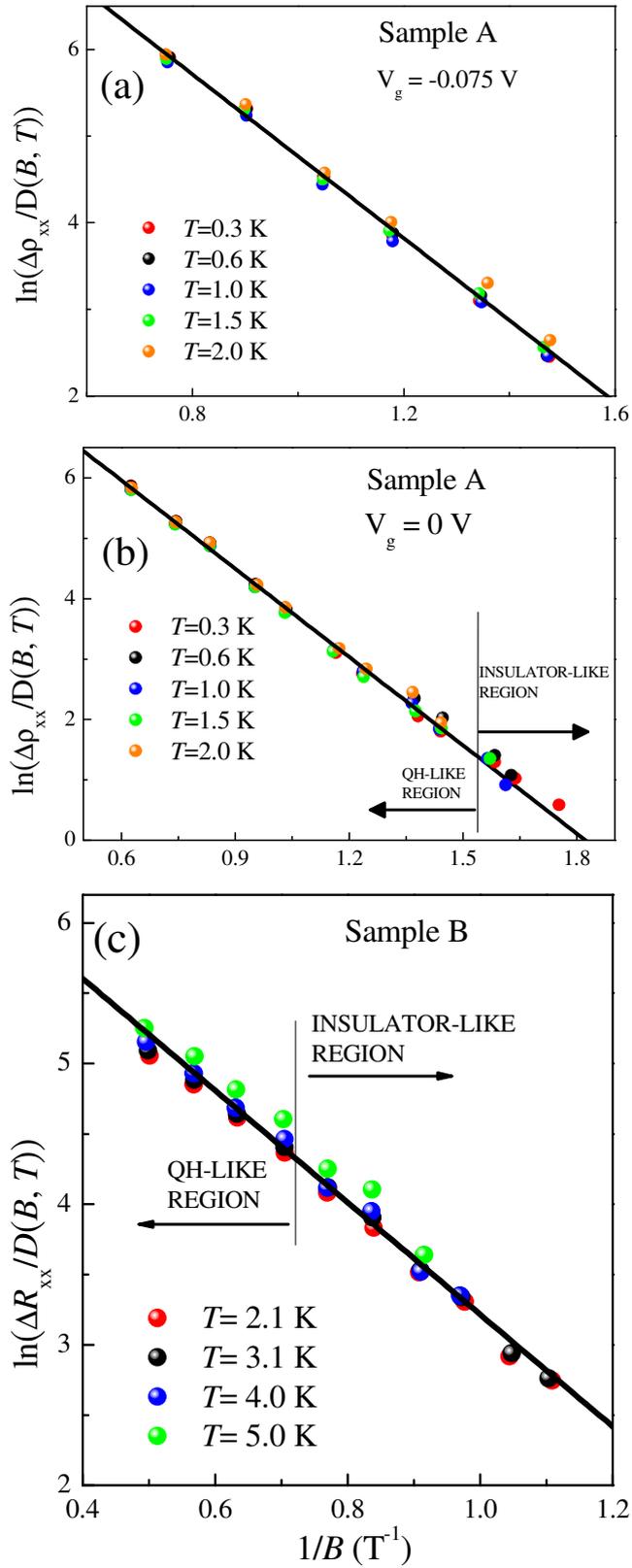

Figure 2

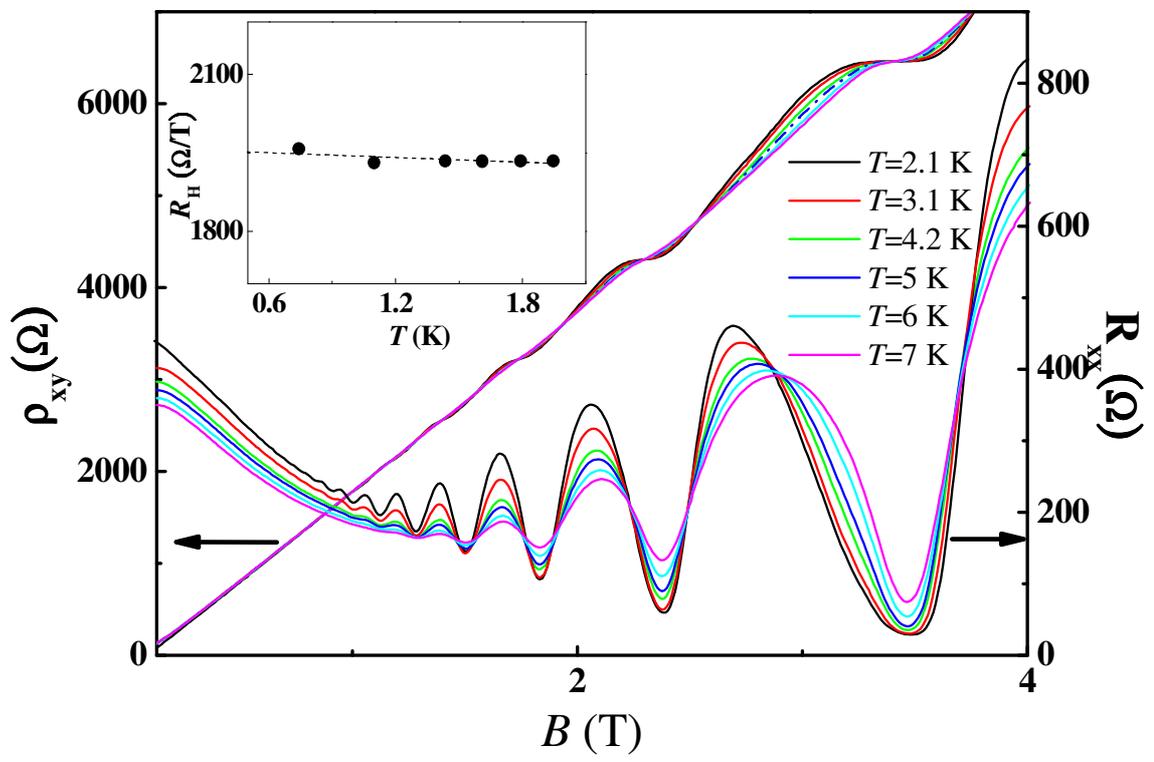

Figure 3

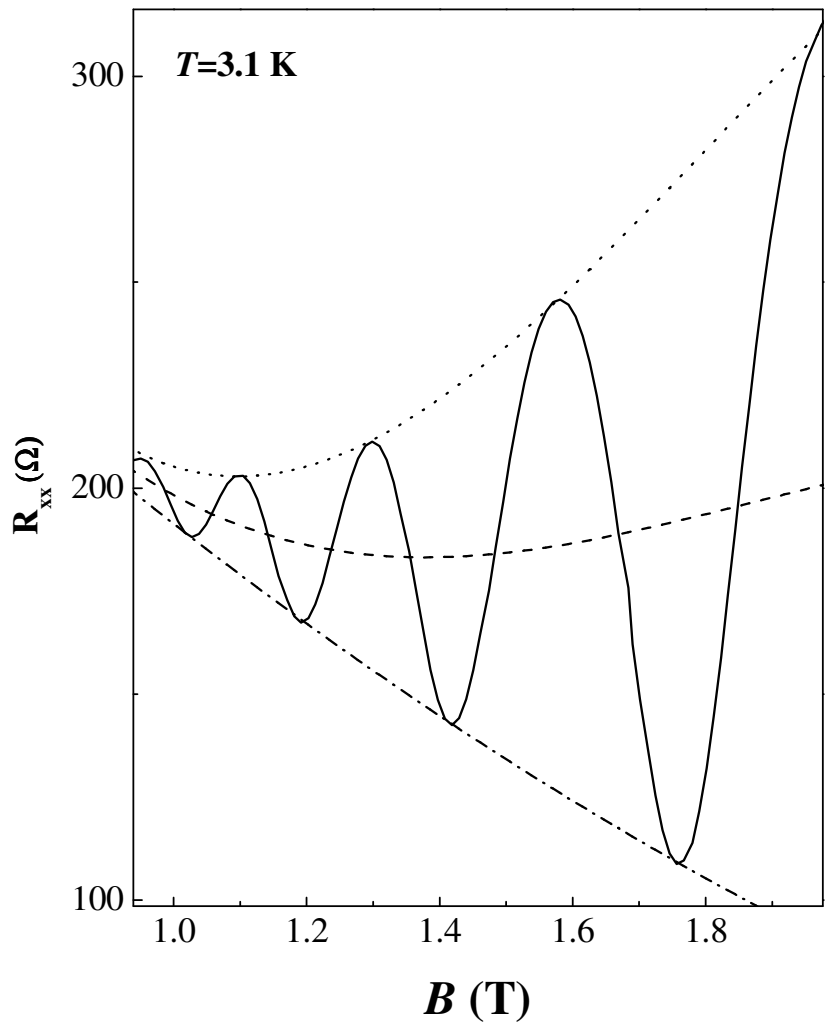

Figure 4

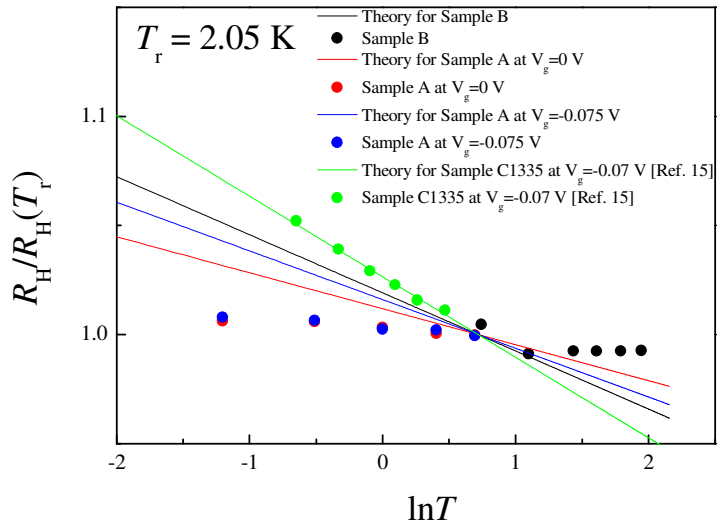

Figure 5